\documentclass[preprint,12pt]{elsarticle}

\usepackage[letterpaper,top=1.2in,bottom=1.5in,left=1.0in,right=1.0in]{geometry}%

\usepackage{float}
\usepackage{amssymb,mathrsfs,amsmath}
\usepackage{MnSymbol,bbding,pifont}
\usepackage{extarrows}
\usepackage{soul}
\usepackage[pdfencoding=auto]{hyperref}
\usepackage{bookmark}
\hypersetup{hidelinks,
	colorlinks=true,
	allcolors=black,
	pdfstartview=Fit,
	breaklinks=true}

\usepackage{color}
\usepackage{courier}
\usepackage{booktabs}
\usepackage{makecell}
\usepackage{multirow}
\usepackage{siunitx}
\usepackage{bm}
\usepackage[figuresright]{rotating}
\usepackage[dvipsnames]{xcolor}
\usepackage[most]{tcolorbox}
\usepackage[toc,page,header]{appendix}
\usepackage{minitoc}
\usepackage{soul}

\usepackage[switch]{lineno}

\usepackage{tabularx}
\usepackage{booktabs}
\usepackage{enumitem}

\newcolumntype{L}{>{\raggedright\arraybackslash}X}
\newcolumntype{C}{>{\centering\arraybackslash}X}

\renewcommand\appendix{\par
    \setcounter{section}{0}
    \setcounter{subsection}{0}
    \gdef\thesection{\Alph{section}}}

{}
{}
{}
{}

\journal{XXXXXXXXXXX}

\begin{document}
\doparttoc 
\faketableofcontents 
\parttoc 

\begin{frontmatter}

\title{Embracing Flow Unsteadiness: A High-Throughput Learning Platform Enables Vortex-Exploiting Bioinspired Propulsion}

\author[a,b]{Fei Han\textsuperscript{$\dagger$}}
\author[b,c]{Xinyu Cui\textsuperscript{$\dagger$}}
\author[b,e]{Zhipeng Wang\textsuperscript{$\dagger$}}
\author[c]{Ning Yang}
\author[b]{Hang Xu}
\author[c]{Haifeng Zhang}
\author[f]{Zhongming Hu}
\author[d]{Jun Wang}
\author[g]{Junfeng Du}
\author[b]{Dixia Fan\textsuperscript{$\ast$}}\ead{fandixia@westlake.edu.cn}

\affiliation[a]{organization={College of Computer Science and Technology},
            addressline={Zhejiang University}, 
            city={Hangzhou},
            country={China},
            }
\affiliation[b]{organization={School of Engineering},
            addressline={Westlake University}, 
            city={Hangzhou},
            country={China},
            }
\affiliation[c]{organization={Institute of Automation},
            addressline={Chinese Academy of Sciences}, 
            city={Beijing},
            country={China},
            }
\affiliation[d]{organization={Department of Computer Science},
            addressline={University College London}, 
            city={London WC1E 6BT},
            country={United Kingdom},
            }
\affiliation[e]{organization={School of Mechanical and Material Engineering},
            addressline={Queen's University}, 
            city={Kingston, ON K7L 3N6},
            country={Canada},
            }
\affiliation[f]{organization={College of Building Engineering},
            addressline={Huanghuai University}, 
            city={Zhumadian},
            country={China},
            }
\affiliation[g]{organization={College of Engineering},
            addressline={Ocean University of China}, 
            city={Qingdao},
            country={China},
            }

\nonumnote{\textsuperscript{$\dagger$} These authors contributed equally to this work.}
\cortext[cor1]{Corresponding author}

\begin{abstract}
Biological swimmers and flyers exploit unsteady vortices for propulsion, whereas engineered vehicles usually suppress them as disturbances. Learning such flow exploitation in machines is difficult because real-fluid interaction data are scarce and unstructured exploration is unstable in high-dimensional, history-dependent flows. Here we present REEF, a co-designed physical-learning framework that integrates SHOAL, an eight-channel high-throughput array for real fluid--structure interaction, with V-STAR, a staged algorithm that converts these interactions into policies through imitation, offline internalization, and online adaptation. Across lift-based, drag-based, and momentum-jet propulsors, REEF expands the attainable force envelope to more than twice that of parameterized search. Particle image velocimetry shows that these gains arise from coordinated vortex formation, growth, and force projection, rather than refinement of a fixed motion-to-force mapping. Force-trained policies transfer zero-shot to free-moving robots whose body motion changes the surrounding flow, suggesting that REEF learns transferable wake-coupling principles for embodied propulsion in unsteady fluids.
\end{abstract}

\end{frontmatter}

\section{Introduction}
\label{sec:intro}

Modern aerial and marine vehicles are built upon design paradigms \cite{panda2021review} that prioritize steady operation, structural rigidity, and functional decoupling, with distinct mechanical components assigned to propulsion, maneuvering, and stabilization in fluid environments \cite{greenblatt2022flow}. Conventional control surfaces, such as fixed wings and propellers, are designed for regimes in which fluid forces are treated as quasisteady responses to instantaneous geometry and inflow changes \cite{abbott2012theory}. Such frameworks have enabled reliable modeling and optimization across diverse engineering systems; however, the flow history is neglected \cite{fossen2011handbook}. Moreover, by treating unsteadiness as a disturbance to be avoided, this assumption forfeits adaptability precisely where strong unsteadiness and nonlinear coupling dominate, the very regimes in which nature excels.

This superior biological performance involves the use of oscillations not only for actuation but also as a precise mechanism for organizing and manipulating flow structures
 \cite{fish2006, taha2020, tack2024fish}. Oscillatory motions generate forces \cite{jones2022} through intertwined mechanisms, including circulation-based lift, separation-induced drag, and added-mass effects. These mechanisms depend critically on how vortices \cite{smits2019} are created, convected, and reencountered over time \cite{liu2024vortices}. Consequently, radically different loads can be produced under identical instantaneous kinematic conditions depending on the time-evolving wake \cite{izraelevitz2014adding}.

Owing to this history dependence and inherent nonlinearity, such systems are intrinsically difficult to control. Minor shifts in phase or frequency can trigger abrupt wake transitions between thrust and drag regimes \cite{andersen2017}, and each control action reshapes the flow conditions that, in turn, govern all future forces \cite{becker2015hydrodynamic}. Consequently, controlling biological propulsion in fluids involves strongly coupled nonlinear, unsteady, and history-dependent dynamics, rendering conventional control design fundamentally challenging.

Existing mainstream approaches, such as parameterized gait design, often approximate fluid response as a mapping from prescribed motion to propulsive force \cite{izraelevitz2014adding, lauder2015fish}. Such formulations have enabled stable and interpretable propulsion in periodic cruising regimes \cite{Chao_2024}, but they become restrictive in agile and vortex-manipulating locomotion \cite{gazzola2012c}. In these regimes, effective control requires high-dimensional and temporally coordinated kinematics: although the motion may retain a gross periodic structure, cycle-to-cycle variations in phase, amplitude, and waveform are essential for regulating the wake \cite{wang2024learn, zhu2025intermittent}. Low-dimensional gait parameterizations therefore confine optimization to a reduced manifold, leaving much of the achievable performance envelope inaccessible. Realizing the full potential of bioinspired propulsion thus requires control paradigms that incorporate temporal reasoning and high-dimensional kinematic representation, allowing the body to actively exploit rather than suppress unsteady fluid dynamics.

Deep reinforcement learning (DRL) offers an appealing route for bioinspired propulsion by replacing explicit fluid modeling with interaction-driven policy discovery \cite{brunton2020}. Yet its application remains constrained by two unresolved bottlenecks: the lack of a structured learning paradigm and the lack of a scalable physical learning environment. First, purely end-to-end approaches often struggle with the high-dimensional, nonlinear, and history-dependent nature of fluid interaction. Although black-box policies have succeeded in canonical flow-control tasks \cite{verma2018efficient, fan2020reinforcement, ren2024, zhou2025reinforcement}, they can become unstable or settle into conservative local optima when extended to agile vortex manipulation. This limitation suggests that effective propulsion control requires more than deeper networks; it requires a physically structured acquisition of motor intelligence.

Biological locomotion provides a guiding analogy for such structure. Motor skills do not emerge from unconstrained trial and error \cite{byrne1998learning}, but through a developmental progression of imitation, internalization, and adaptation. Imitation constrains exploration to physically admissible behaviors \cite{gopnik1993imitation}; repeated practice internalizes experience into empirical priors \cite{gentner2006recursive}; and sensory feedback adapts control to changing conditions \cite{goldstein2003social}. For engineered systems, this biological principle can be operationalized in a more efficient form: costly real-fluid interaction is used to ground a model of the dynamics, while policy refinement can proceed through structured internalization and targeted online adaptation, rather than unconstrained physical trial and error.

The second bottleneck is the absence of a physical learning space where such intelligence can emerge at scale. Simulation-based DRL provides data volume but remains tied to solver assumptions and sim-to-real gaps \cite{wang2024learn, lin2025learning}, whereas physical learning exposes real fluid--structure interaction but is limited by low throughput, hardware complexity, and narrow task coverage \cite{fan2019robotic, kaufmann2023champion}. What is missing is a real, high-throughput environment that captures causal flow history while supporting live closed-loop adaptation.

To address both bottlenecks, we develop \textbf{REEF} (Real-flow Embodied Experimental Framework), a co-designed physical-learning ecosystem for bioinspired propulsion. REEF integrates \textbf{SHOAL} (Synchronized Hydrodynamic Oscillation Array Lab), an eight-channel closed-loop experimental array that generates real fluid--structure interaction data at $1.8 \times 10^4$ trials per day, with \textbf{V-STAR} (vortex-exploiting sequential three-stage adaptive reinforcement learning), a staged algorithm that converts demonstrations into deployable policies through imitation, offline internalization, and online adaptation. SHOAL addresses the environmental bottleneck by providing scalable real-flow interaction, while V-STAR addresses the structural bottleneck by replacing direct end-to-end search with staged motor acquisition.

\begin{figure}[H]
    \centering
    \centerline{\includegraphics[width=0.97\textwidth]{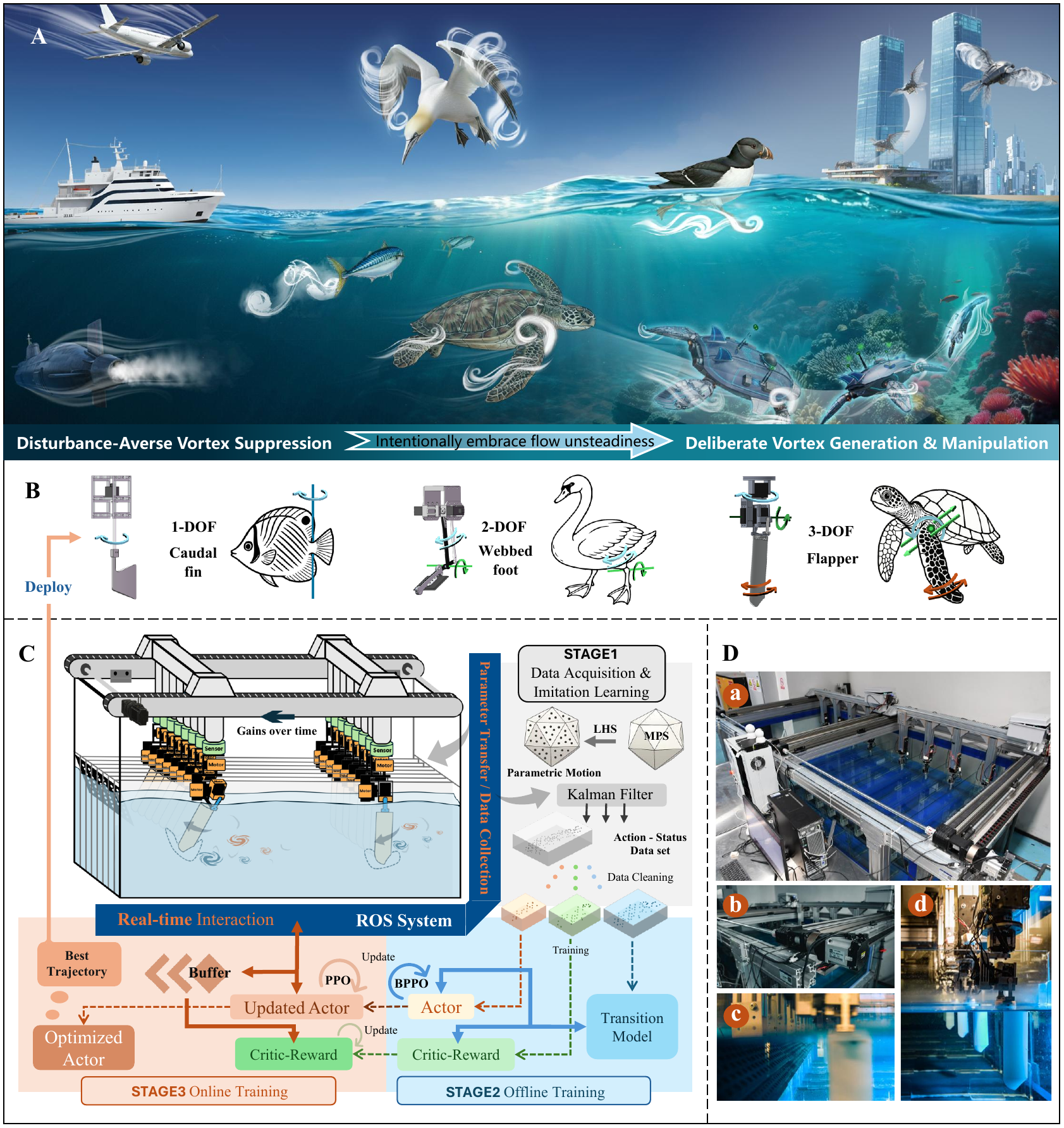}}
    \caption{\textbf{Overview of REEF, a co-designed physical-learning ecosystem integrating the SHOAL experimental array with the V-STAR learning algorithm.} \textbf{(A)} Conventional engineering suppresses vortices to reduce disturbances, whereas biological systems exploit unsteady flows for propulsion; REEF advances toward deliberate vortex generation and manipulation. \textbf{(B)} Three representative propulsors, including a 3-DOF flapper, 2-DOF webbed foot, and 1-DOF caudal fin, span distinct locomotion modes and force-generation regimes. \textbf{(C)} V-STAR converts parameterized trajectories into flow-compliant policies through three-stage algorithm (imitation, offline internalization, and online adaptation). \textbf{(D)} SHOAL enables parallel real-flow learning with an eight-channel towing array, rapid flow reset, synchronized actuation, and real-time six-axis force sensing.}
    \label{fig: main}
\end{figure}

Through this integration, REEF enables propulsion policies to be learned directly within real fluid interactions, moving bioinspired propulsion beyond quasi-steady gait optimization and direct black-box search. Across representative bioinspired propulsors, the learned policies exploit cycle-to-cycle variations in phase, amplitude, and waveform to regulate vortex dynamics and enhance propulsion. In doing so, REEF turns unsteady vortices from disturbances to be suppressed into control resources to be shaped, defining a scalable route toward embodied intelligent fluid robots that do not merely mimic biological kinematics, but learn to co-adapt with the vortical flows they generate (Fig.~\ref{fig: main}A).

\section{REEF: A Co-designed Physical-Learning Ecosystem}
\label{sec:reef}

REEF (Real-flow Embodied Experimental Framework) couples a high-throughput physical experimentation infrastructure, SHOAL, with a staged learning algorithm, V-STAR, and evaluates the resulting framework across three interchangeable bioinspired propulsors (Fig.~\ref{fig: main}). These components address complementary requirements of learning propulsion in unsteady fluids: SHOAL provides scalable real flow interaction data, V-STAR converts these interactions into deployable policies, and the propulsor set tests whether the same framework generalizes across distinct fluid force regimes.

\subsection*{SHOAL: Experimental Array}

SHOAL (Synchronized Hydrodynamic Oscillation Array Lab) consists of eight parallel experimental channels operating in real water rather than simulated flow, preserving the nonlinear, unsteady, and history-dependent wake dynamics that govern bioinspired propulsion (Fig.~\ref{fig: main}D). Each channel measures six-axis hydrodynamic loads in real time, providing feedback for reward evaluation and wake-aware policy refinement. Parallel operation supplies the sample volume needed for high-dimensional motion-force datasets, while in-situ inference enables each channel to update actions from real-time hydrodynamic feedback. Hardware specifications and the distributed control architecture are provided in Methods and Supplementary Note~1; an operational demonstration is shown in Supplementary Video~1.

\subsection*{Bioinspired Propulsor}

Three interchangeable propulsors, a 1-DOF caudal fin, a 2-DOF webbed foot, and a 3-DOF flapper (Fig.~\ref{fig: main}B), span representative biological locomotion modes: fish undulation, waterfowl paddling, and turtle flapping. They also cover distinct fluid-force regimes. The caudal fin emphasizes periodic momentum exchange and added-mass effects; the webbed foot captures drag-dominated paddling through asymmetric power and recovery strokes; and the flapper represents lift-based flapping, where coupled pitching, rolling, and twisting enable circulation control and wake shaping. This progression from one to three actuated degrees of freedom provides a structured test of whether REEF can generalize across propulsion physics.

\subsection*{V-STAR: Three-Staged Learning Algorithm}

V-STAR (vortex-exploiting sequential three-stage adaptive reinforcement learning) decomposes policy acquisition into imitation, internalization, and adaptation (Fig.~\ref{fig: main}C). This staged design addresses three control-side difficulties of bioinspired propulsion: high-dimensional kinematics, strong fluid--body coupling, and quasi-periodic temporal structure. Rather than optimizing a policy end-to-end from random hardware exploration, V-STAR first constrains the search to executable motions, then internalizes the observed fluid--body dynamics offline, and finally adapts the policy through closed-loop interaction with the real wake.

\textbf{Stage I, Imitation.} Large-scale open-loop experiments over a broad kinematic space generate a library of executable trajectories. Imitation learning uses this library to initialize policies within mechanically feasible regions \cite{osa2018algorithmic, nair2018overcoming}. This replaces low-dimensional gait parameters with a high-dimensional policy representation while avoiding the inefficient cold start of end-to-end hardware DRL.

\textbf{Stage II, Internalization.} The accumulated state--action--load data are used to learn the short-horizon consequences of fluid--body interaction without continuous hardware access. A conservative value network \cite{kostrikov2021offline}, a physics-grounded transition model \cite{cao2021choose}, and a sequence-memory module jointly support offline policy improvement under fluid memory. This stage allows the policy to account for recent action--load histories and stabilize optimization in the presence of non-Markovian wake effects.

\textbf{Stage III, Adaptation.} The policy is then refined through real-time closed-loop interaction with the physical wake \cite{ibarz2021train}. Hydrodynamic load feedback corrects residual biases from offline learning and aligns the policy with the evolving motion--wake--load coupling, yielding stable propulsion strategies that remain robust to environmental perturbations.

\section{Results}

To verify that REEF addresses these challenges, we focus on four questions: whether it converges under real fluid--structure interaction and its key components are necessary; whether the learned policies reach performance regimes beyond low-dimensional parametric gait search; whether the framework generalizes across bioinspired propulsion of different fluid mechanics; and whether force-centric policies transfer without retraining to free robotic locomotion. Therefore, we use a 3-DOF flapper as the primary benchmark, then extend the framework to a 2-DOF webbed foot, a 1-DOF caudal fin, and three robotic platforms.

\subsection*{Each component addresses a distinct constraint}

\begin{figure}[H]
    \centerline{\includegraphics[width=1.1\textwidth]{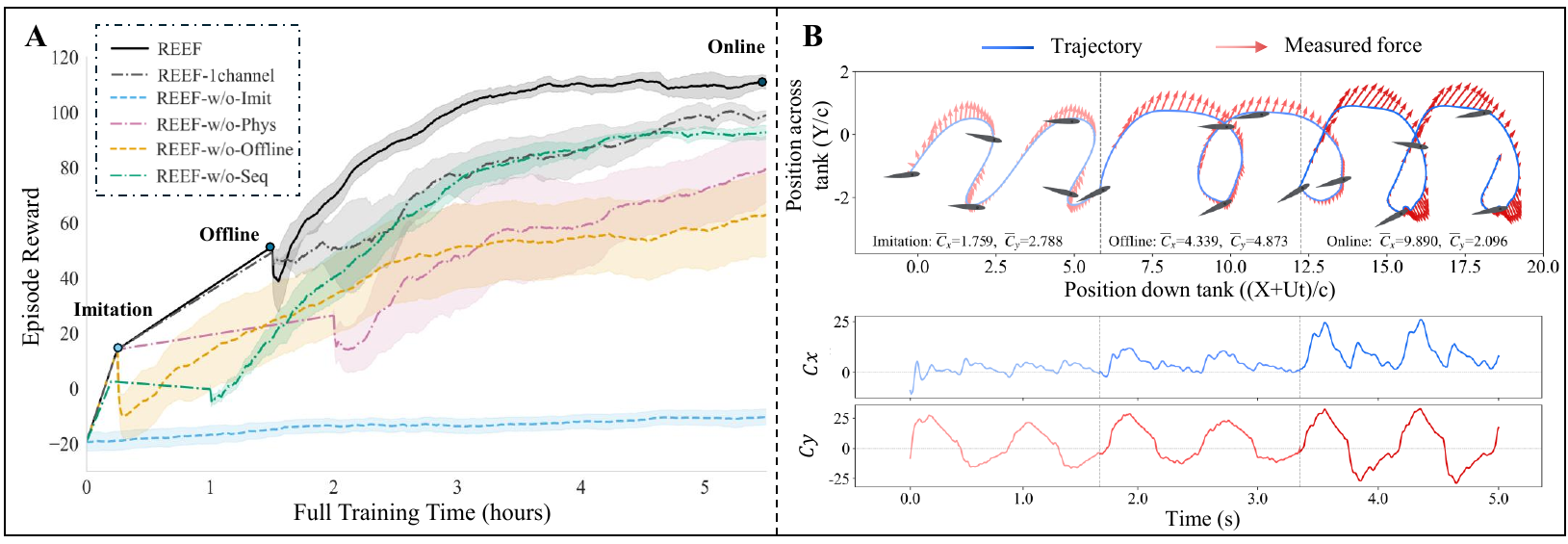}}
    \caption{\textbf{Each component addresses a distinct constraint.} \textbf{(A)} Training reward of the full REEF pipeline and five ablation variants on the 3-DOF flapper benchmark. Curves show smoothed episode reward versus training time; shaded regions denote $\pm 1$ SEM across seeds. Dashed and solid segments indicate pre-training and online learning phases, respectively. Ablation configurations and equivalent baseline algorithms are defined in Table~\ref{tab:ablation_unified}. \textbf{(B)} Kinematic trajectories and hydrodynamic force evolution across the three V-STAR stages. Top: representative flapping-cycle trajectories with instantaneous force vectors and foil orientations. Bottom: thrust ($C_x$) and lateral-force ($C_y$) coefficients along the same cycle, with stage transitions and cycle-averaged thrust $\overline{C_x}$ annotated. Abbreviations: SEM, standard error of the mean.}
    \label{fig: result1}
\end{figure}

Distinguishing framework synergy from redundant complexity requires isolating how each component contributes to learning under real fluid--structure interaction. We therefore performed a controlled ablation study on the 3-DOF flapper benchmark, comparing the full REEF pipeline with five variants that remove or replace one designed element: parallel execution, sequence memory, physics-grounded transition modeling, offline internalization, or imitation-based initialization (Fig.~\ref{fig: result1}A; Table~\ref{tab:ablation_unified}).

\begin{table}[htbp]
\centering
\scriptsize
\caption{\textbf{Ablation performance of REEF variants on the 3-DOF flapper benchmark.}}
\label{tab:ablation_unified}
\begin{tabular*}{\textwidth}{@{\extracolsep{\fill}} l l c c c @{}}
\toprule
\textbf{Variant} & \textbf{Modification (baseline)} & \textbf{Reward} & \textbf{Std} & \textbf{$t_{80\%}$ (h)} \\
\midrule
REEF             & --- (full pipeline)                                & $113.5$           & $\pm\,11.9$           & $2.62$    \\
REEF-1channel    & Single channel                                     & $105.4$           & $\pm\,28.4$           & $4.46$    \\
REEF-w/o-Seq     & Transformer $\rightarrow$ MLP                      & $\phantom{0}99.9$ & $\pm\,\phantom{0}9.9$ & $4.26$    \\
REEF-w/o-Phys    & Physics-grounded $\rightarrow$ statistical (Uni-O4~\cite{lei2023uni})& $\phantom{0}84.1$ & $\pm\,24.3$           & ---       \\
REEF-w/o-Offline & No offline stage (imitation-init PPO)              & $\phantom{0}75.9$ & $\pm\,38.7$           & ---       \\
REEF-w/o-Imit    & No imitation prior (PPO from scratch)              & $\phantom{0-}{-}8.2$ & $\pm\,\phantom{0}4.6$ & ---     \\
\bottomrule
\end{tabular*}
\begin{flushleft}
\tiny \textit{$t_{80\%}$: wall-clock time to reach $80\%$ of REEF's final reward; "---" indicates the variant never reaches this threshold.}
\end{flushleft}
\end{table}

The full REEF pipeline converged smoothly to the highest reward ($113.5 \pm 11.9$ across seeds), whereas each ablation produced a distinct degradation. Removing imitation caused end-to-end learning from scratch to collapse into a static drag-minimizing posture, with no sustained oscillatory propulsion. Removing offline internalization retained the imitation prior but reached only $67\%$ of REEF's final reward and showed substantially larger cross-seed variability, indicating that imitation alone is insufficient for stable hardware learning. Replacing the physics-grounded transition model with a purely statistical baseline plateaued at $74\%$ of REEF's reward, while removing sequence memory reached $88\%$, showing that both physical grounding and temporal action--load history are needed to handle non-Markovian wake dynamics. Finally, reducing SHOAL from eight channels to one preserved much of the final reward ($93\%$) but slowed convergence and increased cross-seed dispersion, confirming that parallel physical experimentation improves not only throughput but also the statistical reliability of hardware-anchored learning.

The cumulative effect of these components is not merely numerical improvement, but a reorganization of the propulsion regime itself (Fig.~\ref{fig: result1}B). Across the three V-STAR stages, the cycle-averaged thrust coefficient increased from $\overline{C_x}=1.759$ after imitation to $4.339$ after offline internalization and $9.890$ after online adaptation, a $5.6\times$ gain over the imitation baseline. This increase was accompanied by a qualitative change in force structure: the single-peak cycle produced by imitation evolved into a double-peak thrust pattern after online adaptation, with instantaneous peaks reaching $\overline{C_x}\approx20$--$25$. Thus, REEF does not simply tune a fixed gait template; it learns a new vortex-coupled propulsion regime, with each component addressing a distinct constraint of physical learning.

\subsection*{Learning expands flapping force envelopes}

A key challenge in bioinspired propulsion is coordinating competing force objectives. Swimming emphasizes thrust, flapping flight requires lift, and agile maneuvering depends on moving continuously between these regimes. Such versatility cannot be captured by a single optimized gait; it requires a policy space capable of organizing multiple force directions across distinct wake-coupled operating points.

\begin{figure}[H]
    \centering
    \centerline{\includegraphics[width=\textwidth]{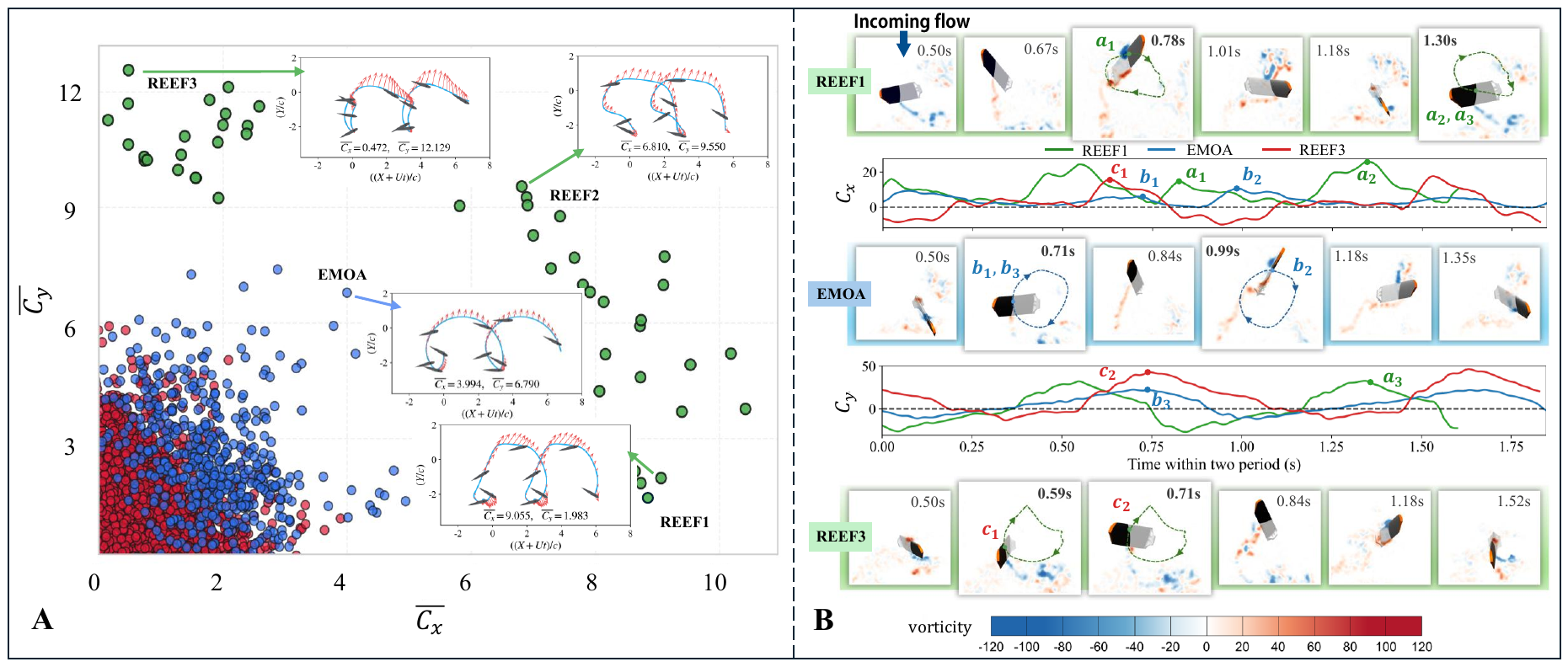}}
    \caption{\textbf{Learned policies access force regimes beyond parameterized search through vortex orchestration.} \textbf{(A)} Thrust--lift Pareto envelope of the 3-DOF flapper. Each point denotes the cycle-averaged force coefficients $(\overline{C_x}, \overline{C_y})$ of one policy from brute-force grid search (BF), evolutionary multi-objective optimization (EMOA) \cite{coello2006evolutionary}, or REEF. Insets show representative REEF policies along the Pareto front, including thrust-dominated REEF1, balanced REEF2, and lift-dominated REEF3, together with the strongest EMOA representative, EMOA4. \textbf{(B)} Simultaneous PIV vorticity fields and force traces for REEF1, EMOA4, and REEF3 over representative cycles. Vorticity snapshots show wing motion and leading-edge vortex (LEV) evolution, while $C_x(t)$ and $C_y(t)$ traces mark the corresponding force peaks. Compared with EMOA4, REEF policies coordinate LEV formation, growth, and force projection with wing orientation, producing expanded thrust--lift performance.}
    \label{fig: result2}
\end{figure}

We therefore evaluated the 3-DOF flapper as a thrust--lift Pareto problem by varying the directional weighting between $\overline{C_x}$ and $\overline{C_y}$ while keeping the learning framework unchanged. REEF was benchmarked against brute-force grid search (BF) and evolutionary multi-objective optimization (EMOA) \cite{coello2006evolutionary}, both searching a 9-dimensional sine-based gait family. REEF substantially expanded the attainable force envelope (Fig.~\ref{fig: result2}A): maximum thrust reached $\overline{C_x}=10.42$, compared with $4.94$ for EMOA and $3.89$ for BF, a $2.11\times$ improvement over the strongest parameterized baseline; maximum lift reached $\overline{C_y}=12.57$, compared with $7.39$ for EMOA.

The resulting policies also departed qualitatively from the parameterized gait family. Three representative policies along the Pareto front exhibited distinct trajectories: a thrust-dominated streamwise flat-cutting motion, a balanced forward--vertical sweep, and a lift-dominated large-amplitude vertical stroke. Thus, changing only the reward direction within the same learning framework produced different force-generation strategies, indicating that REEF accesses wake-coupled control regimes beyond low-dimensional gait search.

To identify the physical origin of this expanded envelope, we examined simultaneous PIV vorticity fields and instantaneous force coefficients for the thrust-specialized policy REEF1 and the strongest EMOA representative, EMOA4 ($\overline{C_x}=3.99$, $\overline{C_y}=6.79$) (Fig.~\ref{fig: result2}B). In REEF1, thrust peaks are synchronized with leading-edge vortex (LEV) evolution: an initial LEV forms at $t=0.78$~s with the secondary thrust peak $a_1$ and is shed by $t=1.01$~s, followed by a second vortex-generation event between $t=1.18$ and $1.30$~s that produces the main thrust peak $a_2$ ($C_x=25.84$, compared with $10.58$ for EMOA4) and the near-maximum lateral-force peak $a_3$. At this moment, the wing orientation projects the LEV-induced force strongly along the rewarded direction.

EMOA4 also generates LEVs, with events near $t=0.71$~s and $t=0.99$~s associated with its force peaks. However, these vortices occur at wing orientations that project only a limited fraction of the vortex-induced force into the target direction. The distinction is therefore not whether LEVs form, but whether their formation, shedding, and force projection are coordinated with the instantaneous three-axis wing state. REEF1 actively times vortex evolution with wing orientation, whereas EMOA4 generates vortices as byproducts of a sine-parameterized motion family.

The lift-specialized policy REEF3 reveals a complementary strategy. A small-attack-angle sweep initiates an LEV, followed by rapid pitch-up that amplifies vortex formation; with the wing then oriented to project the enlarged vortex force laterally, $C_y$ reaches $46.53$, or $2.06\times$ the maximum lateral force produced by EMOA4. Detailed PIV sequences are provided in the Supplementary Materials. Together, these results show that the expanded Pareto envelope arises from active orchestration of LEV formation, growth, shedding, and force projection, rather than finer tuning within a fixed motion-to-force mapping.

\subsection*{Generalization across distinct bio-inspired propulsors}

A framework that learns fluid-force principles rather than platform-specific solutions should generalize across propulsors governed by different mechanics. To test this, we applied the unchanged REEF pipeline to two additional propulsors: a 2-DOF webbed foot representing drag-based paddling and a 1-DOF caudal fin representing momentum-jet undulation.

\begin{figure}[H]
    \centering
    \centerline{\includegraphics[width=\textwidth]{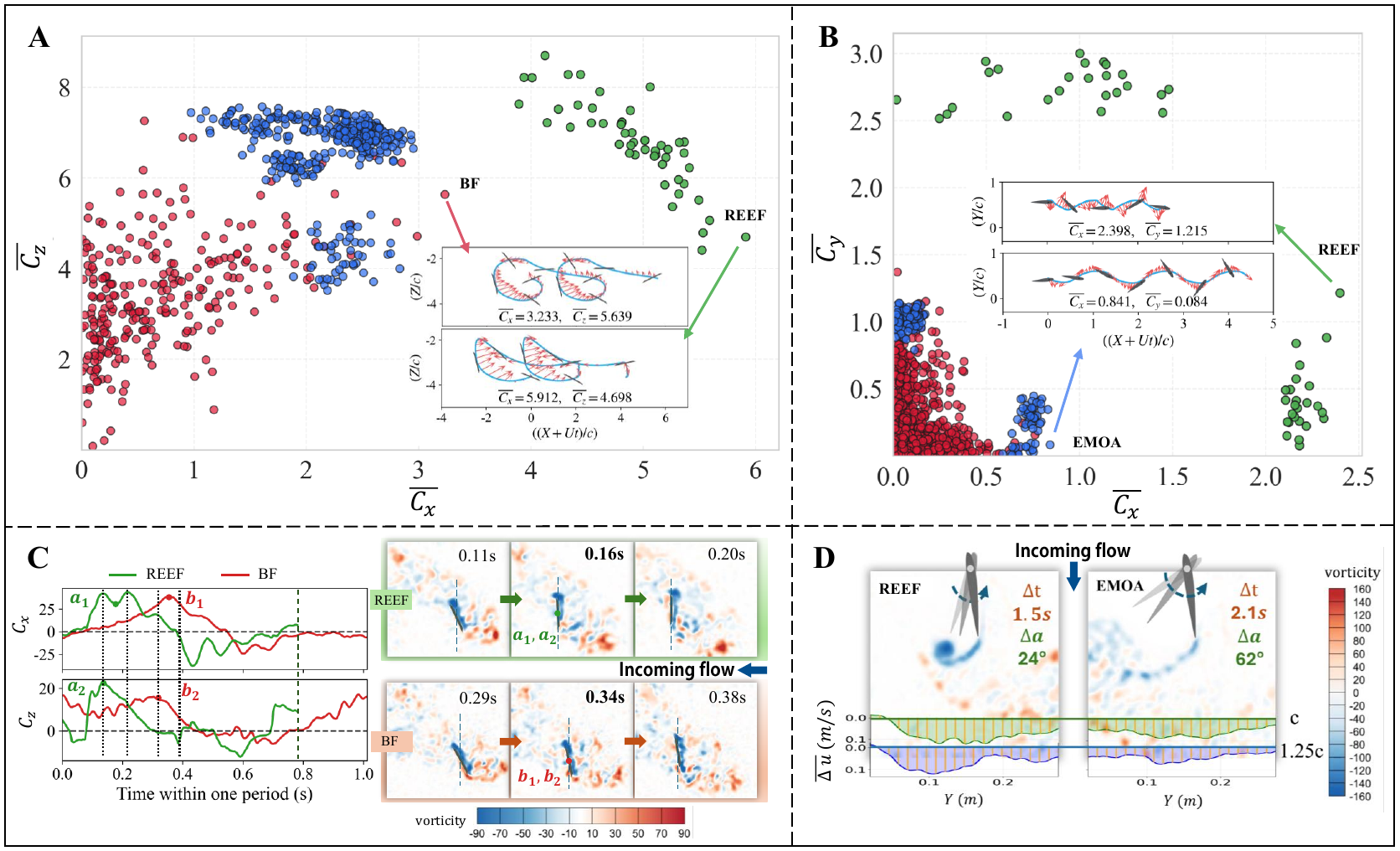}}
    \caption{\textbf{REEF generalizes to drag-based and momentum-jet propulsion through morphology-specific vortex strategies.} \textbf{(A)} Thrust--lift Pareto envelope of the 2-DOF webbed foot. Points denote cycle-averaged force coefficients $(\overline{C_x}, \overline{C_z})$ from BF, EMOA, and REEF; insets show representative trajectories. \textbf{(B)} Thrust--lift Pareto envelope of the 1-DOF caudal fin, with representative REEF and EMOA trajectories. \textbf{(C)} Force traces and PIV vorticity fields for webbed-foot paddling. Snapshots compare REEF at $t=0.11, 0.16, 0.20$~s with BF at $t=0.29, 0.34, 0.38$~s, showing that REEF retains an attached vortex while reorienting the paddle into a thrust-favorable posture. \textbf{(D)} Wake structure and momentum-jet profile for the caudal fin. REEF forms a focused downstream jet, quantified by the time-averaged streamwise velocity increment $\overline{\Delta u}$ along monitoring lines at $c$ and $1.25c$ downstream of the trailing edge. Colorbars show spanwise vorticity; blue and red denote opposite signs.}
    \label{fig: result3}
\end{figure}

\textbf{Webbed-foot paddling: learning drag-based force production.} On the 2-DOF webbed foot, REEF expanded the attainable $(\overline{C_x}, \overline{C_z})$ envelope, increasing maximum thrust to $\overline{C_x}=5.912$, a $2.00\times$ extension over EMOA ($2.952$), while maximum $\overline{C_z}$ reached $8.706$ versus $7.578$ ($1.15\times$) (Fig.~\ref{fig: result3}A). The learned gait also differed qualitatively from the parameterized representative: REEF produced two comparable thrust peaks and a clean vertical-force pulse, whereas the parameterized gait produced a single thrust peak followed by noisy force decay (Fig.~\ref{fig: result3}C, point $b_1$). This indicates that REEF does not merely tune stroke amplitude or phase, but discovers a different drag-based force-generation trajectory.

PIV fields reveal a mechanism different from the flapper case, supporting cross-physics generalization. Instead of LEV orchestration, the webbed foot relies on sustained vortex retention on the upstream paddle face. REEF rotates the paddle from a tilted posture at $t=0.11$~s to a nearly vertical, thrust-favorable orientation by $t=0.16$~s, maintaining both this orientation and the attached vortex through $t=0.20$~s. By contrast, the parameterized representative remains tilted during the stroke ($t=0.29$--$0.38$~s), allowing vortex detachment before effective force projection. Thus, REEF improves drag-based paddling by coordinating paddle reorientation with vortex lifetime.

\textbf{Caudal-fin undulation: learning momentum-jet propulsion.} On the 1-DOF caudal fin, REEF further expanded the force envelope, with maximum $\overline{C_x}=2.398$ versus $0.841$ for EMOA ($2.85\times$) and maximum $\overline{C_y}=3.001$ versus $1.140$ ($2.63\times$) (Fig.~\ref{fig: result3}B). The learned kinematics again departed from the parameterized optimum: REEF converged to a small-amplitude, high-frequency oscillation ($\Delta a=24^\circ$, $\Delta t=1.5$~s), rather than EMOA's large-amplitude, low-frequency motion ($\Delta a=62^\circ$, $\Delta t=2.1$~s). This shift suggests a transition from amplitude-driven sweeping to frequency-dominated momentum exchange.

PIV measurements confirm that the caudal-fin policy exploits focused downstream momentum-jet formation. Along monitoring lines at $c$ and $1.25c$ downstream of the trailing edge, REEF produced sharper velocity increments, with peak $\overline{\Delta u}$ values of $1.23$~m/s and $1.15$~m/s, compared with $0.93$~m/s and $0.78$~m/s for EMOA (Fig.~\ref{fig: result3}D). The learned policy therefore concentrates momentum into a coherent jet rather than dispersing it through large-amplitude wake oscillations.

Across lift-based flapping, drag-based paddling, and momentum-jet undulation, the same framework discovers morphology-specific strategies governed by distinct fluid mechanics: leading-edge vortex orchestration in the 3-DOF flapper, sustained vortex retention through paddle orientation in the 2-DOF webbed foot, and concentrated momentum-jet shaping in the 1-DOF caudal fin. The corresponding peak-thrust extensions over EMOA are $2.11\times$, $2.00\times$, and $2.85\times$, respectively. These results show that REEF does not tune to a single morphology, but learns transferable fluid-force principles that adapt to different propulsor mechanics.

\subsection*{Zero-shot transfer to robotic locomotion}

A final test is whether force-centric policies learned on stationary, instrumented propulsors transfer without retraining to free robotic locomotion. This transfer is non-trivial: SHOAL training occurs under prescribed inflow and fixed-body conditions, whereas a moving robot continuously changes the relative flow encountered by its propulsors. Moreover, the reward contains hydrodynamic force terms but no displacement or velocity objective. We therefore deployed the SHOAL-trained policies zero-shot on three robotic platforms matching the three propulsor families: a quadrupedal flapper robot, a quadrupedal webbed-foot robot, and a caudal-fin module in a circulating flow channel (Fig.~\ref{fig: result4}A--C). Full mechanical and gait-synchronization details are provided in Supplementary Note 5.

Across all platforms, REEF policies produced larger forward displacement than the strongest parameterized representatives in the same amount of time. The flapper robot advanced $1.36$~m versus $0.82$~m over $8$~s ($1.66\times$); the webbed-foot robot advanced $0.71$~m versus $0.51$~m over $7$~s ($1.39\times$); and the caudal-fin module advanced $0.52$~m versus $0.22$~m under the imposed current ($2.36\times$). These gains were consistent across three Pareto-optimal policies and repeated trials. Force histories further show that the characteristic structures learned in SHOAL persist during locomotion, including dense thrust events in the flapper, double-peak paddling in the webbed foot, and high-frequency pulses in the caudal fin, leading to higher cycle-averaged forces across morphologies.

\begin{figure}[H]
    \centering
    \centerline{\includegraphics[width=\textwidth]{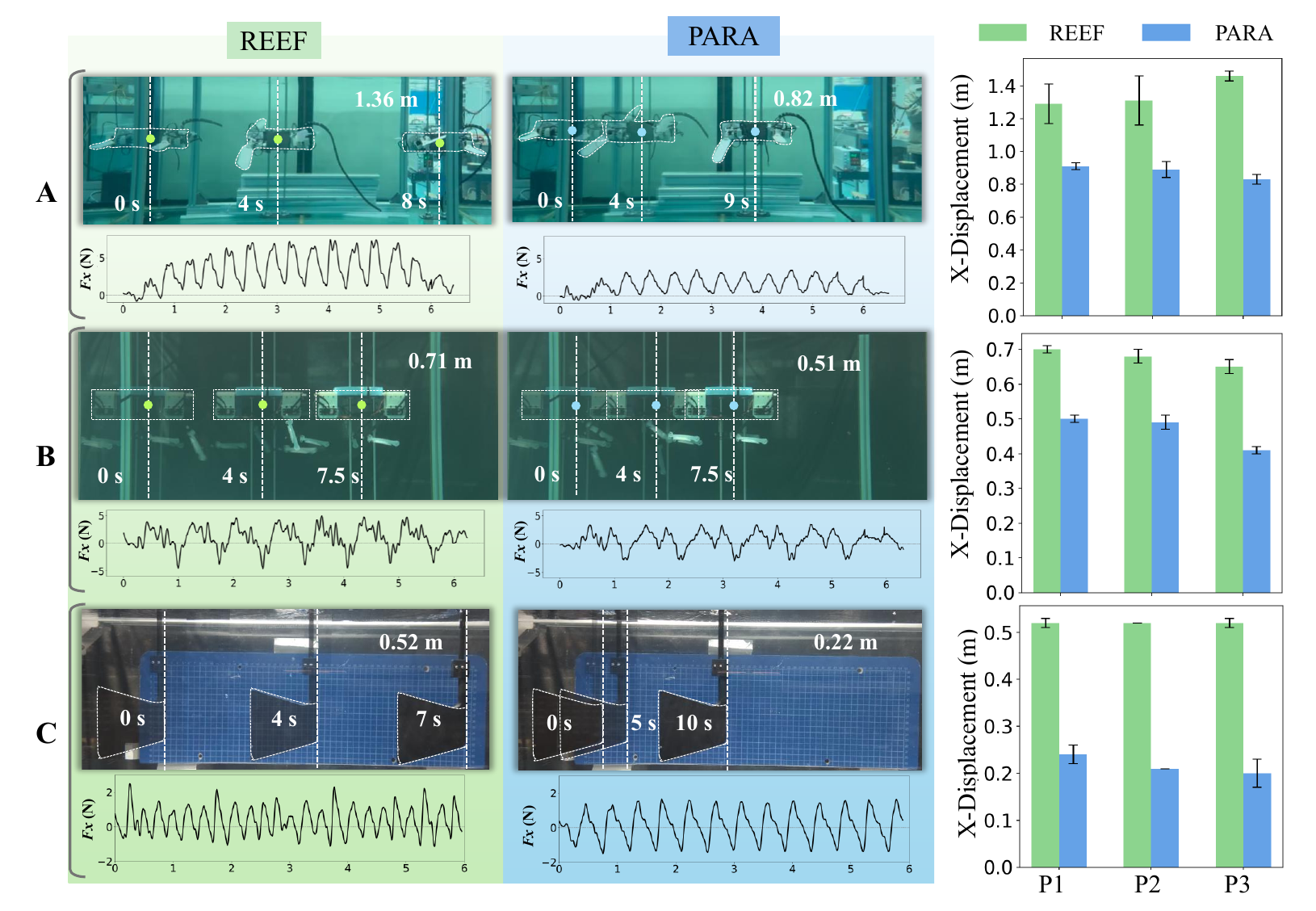}}
    \caption{\textbf{Force-centric policies transfer zero-shot to locomotion across morphologies.} \textbf{(A)} Quadrupedal flapper robot driven by SHOAL-trained REEF policies or the strongest parameterized representative (PARA) from the BF/EMOA pool. Image sequences show body displacement over time, with instantaneous streamwise force $F_x$ below; bars report forward displacement of three REEF policies and corresponding PARA representatives as mean $\pm 1$ s.d. across three trials. \textbf{(B)} Quadrupedal webbed-foot robot under the same protocol. \textbf{(C)} Caudal-fin module tested in a circulating flow channel on near-frictionless air bearings; quadrupedal robots translate along streamwise guide rails with vertical spring compliance. Mechanical and gait-synchronization details are provided in Supplementary Note 5.}
    \label{fig: result4}
\end{figure}

These results indicate that force-centric learning captures motion--wake coordination principles that remain effective under body--flow interaction. However, free locomotion introduces additional coupling: body speed alters wake encounter timing, and shed-vortex strength and orientation can either reinforce or weaken subsequent strokes. Mapping this transfer envelope across body speeds, ambient flows, and whole-body degrees of freedom remains an important direction for future work.

\section{Discussion}

This work reframes bioinspired propulsion as a problem of embodied learning in unsteady fluids. Rather than optimizing predefined kinematic templates, REEF enables propulsors to acquire control strategies directly through sustained physical interaction with the surrounding flow. This capability arises from the co-design of hardware and software: SHOAL provides the high-throughput physical learning space for real fluid--structure interaction, while V-STAR provides the staged learning paradigm that transforms imitation, internalization, and adaptation into deployable propulsion policies. Their integration makes vortical-flow manipulation learnable in hardware for bioinspired propulsors, rather than a mechanism inferred only from biological observation, simplified models, or simulation.

Across three hydrodynamically distinct propulsors, REEF expands the achievable performance envelope by $2.11\times$, $2.00\times$, and $2.85\times$ over conventional parameterized search (EMOA). PIV measurements show that these gains arise not from increased actuation effort, but from organized wake manipulation, including leading-edge vortex orchestration (flapper), sustained vortex retention (webbed foot), concentrated momentum-jet shaping (caudal fin). The $1.35\times$ reward improvement over Uni-O4 further indicates that physical grounding, rather than data volume alone, is critical for learning history-dependent wake dynamics. Finally, the zero-shot transfer of force-trained policies to free-moving robots under unseen flow conditions suggests that REEF captures transferable principles of motion--wake coupling. Together, these results support the central premise of this study: unsteady vortices are not merely disturbances to be suppressed, but control resources that can be shaped.

Several challenges remain before such embodied fluid intelligence becomes general. First, despite their empirical success, the learned policies remain only partially interpretable. V-STAR internalizes history-dependent fluid interactions through experience, but does not yet expose compact rules for when vortices should be generated, retained, or reencountered. A key next step is to embed physical structure more deeply into the learning architecture, through conservation laws, symmetry priors, physics-informed learning \cite{karniadakis2021physics, raissi2020}, sparse system identification \cite{brunton2016discovering}, and operator-based representations \cite{li2020fourier}. This could turn implicit policy knowledge into explicit vortex dynamics, shifting the framework from empirical performance toward explanatory control. Such a transition is essential not only for generalizing design principles across propulsors, scales, and flow regimes, but also for developing machines that illuminate, rather than merely adapt to, the fluid dynamics they exploit \cite{dwivedi2023explainable}.

A second frontier is sensing. The present study relies primarily on rigid-body actuation and global force measurements, which provide only an integrated, low-dimensional summary of the wake. Biological swimmers, by contrast, combine proprioceptive sensing of body deformation with exteroceptive sensing of local flow through pressure, strain, and lateral-line-like systems \cite{triantafyllou2016biomimetic, thandiackal2021emergence, jiao2025sensing}. Embedding spatially distributed and multimodal sensing into future REEF platforms would allow policies to perceive local vortex states, body deformation, and environmental disturbances in real time. Such perception would transform bioinspired robots from force-optimizing actuators into flow-aware embodied agents capable of detecting, harvesting, and responding to environmental energy.

A further step is the transition from single-propulsor force learning to coordinated whole-body locomotion. In a freely moving robot, each fin or limb not only generates force but also reshapes the wake encountered by the rest of the body. Exploiting this coupling will require hierarchical control architectures in which low-level propulsive primitives support high-level posture, trajectory, and task decisions \cite{merel2019hierarchical, lee2020learning}. By jointly regulating body deformation, wake interaction, and thrust production \cite{kim2024wing}, future systems could move beyond fixed operating envelopes toward adaptive motion in confined, cluttered, or turbulent environments.

Beyond individual robots, the same principle may extend to long-term field learning and collective fluid autonomy. As aquatic and aerial agents accumulate experience in real environments, they could progressively refine internal representations of flow, terrain, and body--wake interaction \cite{khetarpal2022towards}. Coupled with high-level reasoning and multi-agent coordination \cite{berlinger2021implicit}, such systems may learn not only how to swim or fly efficiently, but how to exploit wakes, currents, gusts, and confined-channel turbulence as part of their control strategy.

Ultimately, REEF points toward a new generation of embodied intelligent robots in fluids. These systems will not merely imitate biological kinematics or resist the surrounding flow. Instead, they will sense, negotiate, and co-adapt with the fluid environment, achieving agility, robustness, and efficiency through continuous interaction with the unsteady world.

\section{Methods}\label{sec:methods}

\subsection*{Hydrodynamic Force Quantification and Statistical Analysis}

Instantaneous forces in the streamwise direction ($F_x$) and cross-flow directions ($F_y$ lateral, $F_z$ vertical) are nondimensionalized as force coefficients
\begin{equation}
    C_x(t) = \frac{F_x(t)}{\tfrac{1}{2}\rho U^2 S_r}, \qquad C_{y/z}(t) = \frac{F_{y/z}(t)}{\tfrac{1}{2}\rho U^2 S_r},
\end{equation}
where $\rho = 1000$~kg/m$^3$ is the water density, $U = 0.167$~m/s is the towing speed, and $S_r$ is the propulsor reference area: $0.00765$~m$^2$ (flapper), $0.0036$~m$^2$ (webbed foot), and $0.011$~m$^2$ (caudal fin). Cycle-averaged coefficients are computed over $n$ complete motion cycles of period $T$,
\begin{equation}
    \overline{C_x} = \frac{1}{nT}\int_0^{nT} C_x(t)\, dt, \qquad \overline{C_{y/z}} = \frac{1}{nT}\int_0^{nT} C_{y/z}(t)\, dt,
\end{equation}
with $n=5$ used for the Pareto envelopes (Figs.~\ref{fig: result2}A, \ref{fig: result3}A, \ref{fig: result3}B) and $n=3$ used for the momentum-jet profile (Fig.~\ref{fig: result3}D).

All learning curves (Fig.~\ref{fig: result1}A) and ablation rewards in Table~\ref{tab:ablation_unified} report the mean over $N=5$ independent random seeds, averaged over the last 100 episodes; shaded regions show $\pm 1$ standard error of the mean (SEM) and cross-seed dispersion is reported as $\pm 1$ standard deviation (SD). Robotic locomotion bar charts (Fig.~\ref{fig: result4}) show the mean over three independent trials per condition with error bars of $\pm 1$ SD. Pareto envelopes report the union of policies collected across all seeds and reward weightings; no hypothesis testing is applied to envelope extents because the comparison concerns attainable maxima of the policy distribution rather than its expected value. The momentum-jet velocity profiles report time-averaged values without uncertainty bands because each profile is itself a temporal average across three consecutive strokes.

\subsection*{SHOAL: High-Throughput Experimental Array}

SHOAL is the experimental backbone of REEF. The towing tank measures 3.2~m $\times$ 1.5~m $\times$ 1.0~m and is partitioned into eight hydrodynamically separated sub-tanks by 12~mm tempered-glass dividers, which both isolate cross-channel wake interference and stiffen the lateral walls. The towing motors and actuation carriages are mounted on a secondary frame mechanically decoupled from the sub-tank frame through elastomeric isolators, attenuating motor-induced vibration at the six-axis force/torque sensors. Each sub-tank contains a three-layer carbon-fiber wave-damping plate at its downstream boundary, with an automated quiescent interval enforced between consecutive trials. Each channel carries a six-axis force/torque sensor with force resolution $0.25$~N for $F_x, F_y$ and $0.5$~N for $F_z$, torque resolution $0.01$~N$\cdot$m, and a 200~Hz sampling rate. The eight channels are coordinated through a deterministic ROS-based control architecture and together sustain a throughput of $1.8 \times 10^4$ autonomous trials per day, with in-situ policy inference and updates executed on an onboard NVIDIA RTX A4000 GPU. Channel-to-channel consistency was verified using Pearson correlation coefficient (PCC) for waveform shape and normalized root-mean-square error (NRMSE) for amplitude deviation: intra-channel PCC exceeded $0.95$ across three independent repeats and cross-channel NRMSE remained below $12.5\%$ across the eight channels. Full hardware specifications, distributed control topology, and channel-validation results are documented in Supplementary Notes~1--2.

\subsection*{Bio-Inspired Propulsors}

Three propulsors were designed to span lift-based, drag-based, and momentum-jet propulsion regimes (Supplementary Note~2). The \textbf{3-DOF flapper} is modeled on a NACA~0016 symmetric airfoil with a chord of 45~mm and a span of 170~mm (aspect ratio 3.78); three orthogonal rotational axes intersect at a single virtual hinge, providing decoupled flipping, plunging, and sweeping motions each constrained to $[-30^\circ, +30^\circ]$. The \textbf{2-DOF webbed foot} is a trapezoidal membrane (base $60\times 60$~mm, top $40\times 40$~mm, thickness 2~mm) actuated by hip ($\theta_h \in [-45^\circ, +45^\circ]$) and ankle ($\theta_a \in [-60^\circ, +60^\circ]$) rotations. The \textbf{1-DOF caudal fin} is based on a NACA~0012 airfoil with a chord of 100~mm and a span of 110~mm, actuated by a single rotational axis constrained to $[-30^\circ, +30^\circ]$. All actuation is provided by RS-485 bus servos rated at $50$~kg$\cdot$cm with $1^\circ$ positional resolution and $330$~Hz maximum control rate; all structural frames are CNC-machined aluminum alloy.

\subsection*{V-STAR: Three-Stage Learning Algorithm}

\textbf{Overall formulation.} Bioinspired propulsion is formulated as a partially observable Markov decision process (POMDP)~\cite{spaan2012partially} $\langle\mathcal{S}, \mathcal{A}, \Omega, \mathcal{O}, P, R, \gamma\rangle$, in which the latent state $s\in\mathcal{S}$ contains unobservable flow structures, the action $a\in\mathcal{A}$ specifies joint commands, and the observation $o\in\mathcal{O}$ comprises propulsor pose, velocity, and six-axis force/torque measurements. The policy $\pi_\theta$ maximizes the expected discounted return
\begin{equation}
J(\pi_\theta) = \mathbb{E}_{\tau\sim\pi_\theta}\!\left[\sum_{t=0}^{\infty}\gamma^t R(s_t,a_t)\right].
\end{equation}
The step reward is restricted to hydrodynamic force terms and a deadlock penalty:
\begin{equation}
r(t) = a F_x(t) + b F_{y/z}(t) - \iota_t,
\label{eq:reward}
\end{equation}
where $F_x(t)$ and $F_{y/z}(t)$ are the step-averaged thrust and lateral force components, $(a,b)$ are directional weights selected to bias the policy toward thrust, lift, or a balanced mode (per-propulsor values in Supplementary Note~3), and $\iota_t$ is a deadlock penalty activated when joint positional error persists across multiple consecutive steps (formulation in Supplementary Note~3). Equation~\eqref{eq:reward} contains no displacement or velocity term: the policy is trained entirely from hydrodynamic forces measured on a stationary propulsor. This force-only design provides the basis for the embodied-transfer experiments in Fig.~\ref{fig: result4}.

V-STAR is organized into three stages that match the imitation--internalization--adaptation structure of biological motor learning (Fig.~\ref{fig: result1}A). Complete network architectures (Supplementary Note~3 Table~1), training hyperparameters (Supplementary Note~3 Table~2), and the MBPPO pseudocode (Supplementary Note~3 Algorithm~1) are deferred to the supplement.

\textbf{Stage 1: Bootstrapped imitation initialization.}
Stage 1 establishes a feasible behavioral prior over the propulsor's kinematic space. The parametric search space is parameterized by sinusoidal primitives,
\begin{equation}
\theta^i(t) = A_{\theta^i}\sin(2\pi f t + \phi^i) + \theta^i_0,
\end{equation}
where $A_{\theta^i}$, $\phi^i$, $\theta^i_0$ are the amplitude, phase, and offset of the $i$-th degree of freedom, and $f$ is a shared frequency. Latin hypercube sampling (LHS) over this nine-dimensional space on SHOAL produced raw trajectory pools of $6.0 \times 10^4$, $4.0 \times 10^4$, and $2.4 \times 10^4$ trajectories for the flapper, webbed foot, and caudal fin respectively, each trajectory comprising approximately 120 decision steps. The raw pool was curated into two datasets with distinct roles: an expert dataset $\mathcal{D}_{\text{exp}}$ containing non-dominated trajectories on the $(\overline{C_x},\overline{C_{y/z}})$ Pareto front, and a transition dataset $\mathcal{D}_{\text{trans}}$ obtained by density-based sparsification with looser sparsification near the Pareto frontier. $\mathcal{D}_{\text{exp}}$ provides behavioral priors for policy initialization; $\mathcal{D}_{\text{trans}}$ trains the physical transition model used in Stage~2. The EMOA baseline reported in Figs.~\ref{fig: result2}, \ref{fig: result3} is obtained by extending the Pareto frontier of this same sinusoidal family using a Random-Forest-assisted evolutionary multi-objective optimization (EMOA)  active-learning loop (Supplementary Note~3).

To avoid mode averaging across thrust-, lift-, and balance-biased experts, $\mathcal{D}_{\text{exp}}$ was partitioned into three subsets and 12 policies were initialized in parallel (four per subset) via bootstrapped imitation learning, each trained by behavior cloning on its assigned subset. This yielded 12 diverse but physically consistent behavioral priors covering the Pareto front.

\textbf{Stage 2: Model-assisted offline internalization.}
Stage 2 improves the 12 imitation-initialized policies without further physical interaction, using a learned physical model of the propulsor--fluid system as the surrogate environment for policy updates.

A dynamics model $\mathcal{M}_\phi$ predicts the next observation given a history of the past 20 observation--action pairs. The encoder is a Galerkin transformer~\cite{cao2021choose}, chosen to capture non-Markovian dependencies in the action--wake history; the decoder is a Fourier neural operator (FNO)~\cite{li2020fourier}, chosen to represent the propulsor--fluid mapping in the spectral domain where vortex-shedding dynamics are quasi-periodic. The model is trained on $\mathcal{D}_{\text{trans}}$ by minimizing
\begin{equation}
\mathcal{L}_{\text{dyn}}(\phi) = \mathbb{E}_{(h,a,o')\sim\mathcal{D}_{\text{trans}}}\!\left[\|o' - \hat{o}'\|_2^2\right],
\end{equation}
A state-value function $V_\psi$ is trained on $\mathcal{D}_{\text{exp}}$ via implicit Q-learning (IQL)~\cite{kostrikov2021offline} with expectile $\tau = 0.9$, providing a conservative but optimistic anchor for offline policy improvement without querying out-of-distribution actions (loss formulation in Supplementary Note~3).

To avoid both the out-of-distribution extrapolation problem of purely $Q$-based offline updates and the long-horizon divergence of pure model rollouts, the advantage estimate is localized to a one-step physical prediction while preserving long-horizon performance information through $V_\psi$:
\begin{equation}
\hat{A}(h,a) = \operatorname{symlog}(\hat{r}(h,a)) + \gamma\, V_\psi(\hat{h}') - V_\psi(h),
\end{equation}
where $\hat{h}'$ is the next history sequence generated by $\mathcal{M}_\phi$, $\hat{r}$ is the model-predicted immediate reward, and the symlog transform stabilizes occasional reward spikes induced by transient vortex events. This estimate is used inside a behavioral proximal policy optimization (BPPO)~\cite{zhuang2023behavior} clipped objective with $\epsilon = 0.2$ and discount $\gamma = 0.99$; the reference policy is updated periodically by selecting the policy with the higher mean symlog-reward across $N=8$ parallel in-silico rollouts of horizon $H=80$ on $\mathcal{M}_\phi$ (full BPPO objective and reference-update criterion in Supplementary Note~3). This combination defines model-assisted BPPO (MBPPO), the optimizer used inside V-STAR's Stage~2. MBPPO is not a separate framework: it is the offline-stage optimizer of V-STAR, which is the learning algorithm operating on the SHOAL infrastructure within the overall REEF design. Per-component ablations isolating the contributions of the Galerkin--FNO dynamics model, model-based advantage estimation, and model-based reference selection are reported in Supplementary Note~3.

\textbf{Stage 3: Online adaptation on SHOAL.}
The best offline policy is selected by physical evaluation on SHOAL and refined online using proximal policy optimization (PPO)~\cite{schulman2017proximal} with the same reward function as Stage~2. Stage~3 compensates for residual discrepancies between $\mathcal{M}_\phi$ and the physical system (sensor noise, actuator hysteresis, unresolved flow effects) rather than re-learning the dominant hydrodynamic structure. PPO objective, training schedule, and deadlock-penalty formulation are in Supplementary Note~3.

\subsection*{Particle Image Velocimetry}

Two-dimensional particle image velocimetry (PIV) was used to resolve the unsteady vortex fields underlying the kinematic strategies reported in Figs.~\ref{fig: result2}, \ref{fig: result3}. All measurements were conducted at the same free-stream condition as the training experiments, $U_\infty = 0.167$~m/s. The working fluid was seeded with $10$~$\mu$m neutrally buoyant hollow glass spheres; illumination was provided by a 10~W, 532~nm continuous laser sheet of approximately 1.5~mm thickness. Images were captured with a Photron FASTCAM Mini UX50 high-speed camera fitted with a 50~mm $f/1.4$ Nikon prime lens, at 1000~fps for the flapper and 640~fps for the webbed foot and caudal fin. Vector fields were computed in PIVlab using multi-pass FFT cross-correlation (64$\times$64 then 32$\times$32 pixel interrogation windows, 50\% overlap) with three-point Gaussian sub-pixel interpolation; spurious vectors were rejected using an $8\sigma$ standard-deviation filter combined with a local median filter and replaced via interpolation. Out-of-plane vorticity was computed from the resulting velocity fields and is reported in units of $\mathrm{s^{-1}}$. Full optical configuration, per-mechanism acquisition windows, and trigger timing are documented in Supplementary Note~4.

\subsection*{Robotic Verification Platforms}

Three robotic platforms were built to test transfer of force-trained policies to embodied locomotion (Fig.~\ref{fig: result4}; Supplementary Note~5). The \textbf{quadrupedal flapper robot} integrates four 3-DOF flapper modules around a centrally buoyant frame ($216 \times 100 \times 92$~mm, neutrally buoyant). The \textbf{quadrupedal webbed-foot robot} uses four 2-DOF webbed-foot modules in an expanded chassis ($278 \times 190 \times 114$~mm, neutrally buoyant). Both quadrupedal platforms are suspended from a dual-shaft $x$-axis guide rail with linear ball bearings, allowing free streamwise translation; an orthogonal $z$-axis guide system with a $49$~N/m extension spring provides vertical compliance against unsteady hydrodynamic loads, with equivalent $x$-axis damping of $2.95/4.4$~N$\cdot$s/m (kinetic/static). The \textbf{caudal-fin verification platform} mounts a single 1-DOF caudal-fin module on two parallel guide rods supported by air bearings pressurized at $2$~bar, reducing $x$-axis damping to approximately $0.8$~N$\cdot$s/m so that streamwise motion is driven almost exclusively by hydrodynamic force; it operates in a circulating water channel at an imposed free-stream velocity of $0.22$~m/s.

For the quadrupedal platforms, the diagonal-gait test sequence proceeds through static, acceleration, stable-cruising, and coast-to-stop phases. The front-left and back-right limbs initiate the gait at $t=0$; the front-right and back-left limbs are held stationary for a delay $\tau$ before joining, with $\tau = 0.5T$ for the flapper robot (anti-phase delay maximizing lateral-force cancellation) and $\tau = 0.35T$ for the webbed-foot robot (chosen to minimize total-force fluctuation while preventing mechanical collision between front and hind limbs during their extended stroke trajectories). For DRL policies, the dominant actuation period $T$ is extracted via fast Fourier transform of the control signal; for parameterized baselines, $T$ is intrinsically prescribed. The caudal-fin platform is tested without multi-limb constraints by executing the policy from the initial position until the carriage reaches the end of the air-bearing rail. For each robot, three Pareto-optimal DRL policies (P1, P2, P3) and three corresponding parameterized representatives were each evaluated across three independent trials. Full mechanical specifications, buoyancy calibration, and gait-synchronization formulation are in Supplementary Note~5.

\section*{Data availability}
The datasets generated and/or analyzed during the current study will be available on GitHub upon publication.

\section*{Code availability}
The custom source code used for the reinforcement learning framework and distributed control will be available on GitHub upon publication.

\section*{Competing interests}
The authors declare that they have no competing interests.

\section*{Supplementary information}

\noindent\textbf{Supplementary Note 1.} SHOAL physical infrastructure and distributed control architecture.
\vspace{0.5em}

\noindent\textbf{Supplementary Note 2.} Bioinspired propulsor design and cross-channel experimental validation.
\vspace{0.5em}

\noindent\textbf{Supplementary Note 3.} V-STAR three-stage learning algorithm: network architectures, training procedures, and ablation studies.
\vspace{0.5em}

\noindent\textbf{Supplementary Note 4.} Particle image velocimetry configuration and flow-field analysis.
\vspace{0.5em}

\noindent\textbf{Supplementary Note 5.} Robotic verification platforms and gait-synchronization protocols.
\vspace{0.5em}

\noindent\textbf{Supplementary Video 1.} Operational paradigm of the BPL-Farm 8-channel aquatic platform.

\vspace{0.5em}
\noindent\textbf{Supplementary Video 2.} Vortex dynamics observation of bioinspired propulsion under matching flow conditions.

\vspace{0.5em}
\noindent\textbf{Supplementary Video 3.} Zero-shot transfer of learned force-centric policies to three robotic platforms.

\clearpage

 \bibliographystyle{elsarticle-num} 
 \bibliography{cas-refs.bib}


\end{document}